\documentclass[12pt,preprint]{aastex}
\usepackage {graphicx}
\usepackage{aastexug}

\shorttitle{SCREW INSTABILITY OF MAGNETIC FIELD AND GAMMA-RAY BURSTS
IN TYPE IB/C SUPERNOVAE} \shortauthors{D. X. Wang et al.}

\begin{document}

\title{SCREW INSTABILITY OF MAGNETIC FIELD AND GAMMA-RAY BURSTS IN
TYPE IB/C SUPERNOVAE}

\author{Ding-Xiong Wang\altaffilmark{1},  Wei-Hua Lei  and  Yong-Chun Ye}
\affil{Department of Physics, Huazhong University of Science and
Technology,Wuhan, 430074, P. R. China}

\altaffiltext{1}{Send offprint requests to: D. X. Wang
(dxwang@hust.edu.cn)}

\begin{abstract}

A toy model for gamma-ray burst supernovae (GRB-SNe) is discussed by
considering the effects of screw instability of magnetic field in
black hole (BH) magnetosphere. The screw instability in the
Blandford-Znajek (BZ) process (henceforth SIBZ) can coexist with the
screw instability in the magnetic coupling (MC) process (henceforth
SIMC). It turns out that both SIBZ and SIMC occur inevitably,
provided that the following parameters are greater than some
critical values, i.e., (i) the BH spin, (ii) the power-law index
describing the magnetic field at the disk, and (iii) the vertical
height of the astrophysical load above the equatorial plane of the
rotating BH. The features of several GRBs are well fitted. In our
model the durations of the long GRBs depend on the evolve time of
the half-opening angle. A small fraction of energy is extracted from
the BH via the BZ process to power a GRB, while a large fraction of
energy is extracted from the BH via the MC process to power an
associated supernova. In addition, the variability time scales of
tens of msec in the light curves of the GRBs are fitted by two
successive flares due to SIBZ.

\end{abstract}

\keywords{ accretion, accretion disks -- black hole physics -- gamma
rays: bursts-supernovae: general --- magnetic fields ---
instability}

\section{INTRODUCTION}

Recently, observations and theoretical considerations have linked
long-duration GRBs with ultra-bright Type Ib/c supernovae (SNe;
Galama et al. 1998, 2000; Bloom et al. 1999). The first candidate
was provided by SN 1998bw and GRB 980425, and the recent HETE-II
burst GRB 030329 has greatly enhanced the confidence in this
association (Stanek et al. 2003; Hjorth et al. 2003). Extremely high
energy released in very short time scale suggests that GRBs involve
the formation of a black hole (BH) via a catastrophic stellar
collapse event or possibly a neutron star merger, implying that an
inner engine could be built on an accreting BH.

Among a variety of mechanisms of powering GRBs the Blandford-Znajek
(BZ) process (Blandford {\&} Znajek, 1977) has its unique advantage
in providing ``clean'' (free of baryonic contamination) energy by
extracting rotating energy from a BH and transferring it in the form
of Poynting flow in the outgoing energy flux (Lee et al. 2000,
hereafter Lee00; Li 2000c, hereafter Li00).

Not long ago Brown et al. (2000, hereafter B00) worked out a
specific scenario for a GRB-SN connection. They argued that the GRB
is powered by the BZ process, and the SN is powered by the magnetic
coupling (MC) process, which is regarded as one of the variants of
the BZ process (Blandford 1999; van Putten 1999; Li 2000b, 2002;
Wang, Xiao {\&} Lei 2002, hereafter W02). It is shown in B00 that
about $10^{53}ergs$ are available to power both a GRB and a SN.
However, they failed to distinguish the fractions of the energy for
these two objects.

More recently, van Putten and his collaborators (van Putten 2001;
van Putten {\&} Levinson 2003, hereafter P03) worked out a poloidal
topology for the open and closed magnetic field lines, in which the
separatrix on the horizon is defined by a finite half-opening angle.
The duration of a GRB is set by the lifetime of the rapid spin of
the BH. It is found that GRBs and SNe are powered by a small
fraction of the BH spin energy. This result is consistent with
observations, i.e., duration of GRBs of tens of seconds, true GRB
energies distributed around $5\times 10^{50}ergs$ (Frail et al.
2001), and aspherical SNe kinetic energies of $2\times 10^{51}ergs$
(Hoflich et al. 1999).

Very recently, Lei et al. (2005, hereafter Lei05) proposed a
scenario for GRBs in type Ib/c SNe invoking the coexistence of the
BZ and MC processes. In Lei05 the GRB is powered by the BZ process,
and the associated SN is powered by the MC process. The overall time
scale of the GRB is fitted by the duration of the open magnetic flux
on the horizon.

Besides the features of high energy released in very short durations
most GRBs are highly variable, showing very rapid variations in flux
on a time scale much shorter than the overall duration of the burst.
Variability on a time scale of milliseconds has been observed in
some long bursts (Norris et al. 1996; McBreen \textit{et al.}, 2001;
Nakar and Piran, 2002). Unfortunately, the origin of the variations
in the fluxes of GRBs remains unclear. In this paper we intend to
discuss the mechanism for producing the variations in the fluxes of
GRBs by virtue of the screw instability in BH magnetosphere.

It is well known that the magnetic field configurations with both
poloidal and toroidal components can be screw-unstable. According to
the Kruskal-Shafranov criterion the screw instability will occur, if
the toroidal magnetic field becomes so strong that the magnetic
field line turns around itself once or more (Kadomtsev 1966; Bateman
1978).

Some authors have addressed the screw instabilities in BH
magnetosphere. Gruzinov (1999) argued that the magnetic field with a
bunch of closed field lines connecting a Kerr BH with a disk can be
screw-unstable, resulting in the release of magnetic energy with the
flares at the disk. Li (2000a) discussed the screw instability of
the magnetic field in the BZ process, leading to a stringent upper
bound to the BZ power. Wang et al. (2004, hereafter W04) studied the
screw instability in the MC process. They concluded that this
instability could occur at some place away from the inner edge of
the disk, provided that the BH spin $a_ * $ and the power-law index
$n$ for the variation of the magnetic field on the disk are greater
than some critical values.

In this paper we attempt to combine the screw instability of the
magnetic field with the coexistence of the BZ and MC processes. To
facilitate the description henceforth we refer to the screw
instability of the magnetic field occurring in the BZ and MC
processes as SIBZ and SIMC, respectively. It is shown that both SIBZ
and SIMC can occur, provided that the following parameters are
greater than some critical values: (\ref{eq1}) the BH spin,
(\ref{eq2}) the power-law index describing the variation of the
magnetic field at the disk, and (\ref{eq3}) the vertical height of
the astrophysical load above the equatorial plane of the Kerr BH.

The features of several GRB-SNe are well fitted in our model.
(\ref{eq1}) The overall duration of the GRBs is fitted by the
evolution of the half-opening angles. (\ref{eq2}) The true energies
of several GRBs are fitted by the energy extracted in the BZ
process, and the energies of associated SNe are fitted by the energy
transferred in the MC process. (\ref{eq3}) The variability time
scales of tens of msec in the light curves of several GRBs are
fitted by two successive flares due to SIBZ.

This paper is organized as follows. In $\S$ 2 we derived a criterion
of SIBZ based on the Kruskal-Shafranov criterion and some simplified
assumptions on the remote load. In $\S$ 3 we discuss the time scale
and energy extraction from a Kerr BH in the context of the suspended
accretion state. In $\S$ 4 we propose a scenario for the origin of
the variation in the light curves of GRBs based on the flares
arising from SIBZ. Finally, in $\S$ 5, we summarize the main results
and discuss some issues related to our model. Throughout this paper
the geometric units $G = c = 1$ are used.

\section{SCREW INSTABILITY IN BH MAGNETOSPHERE}

In W04 the criterion of SIMC is derived based on the following
points: (\ref{eq1}) the Kruskal-Shafranov criterion, (\ref{eq2}) the
mapping relation between the angular coordinate on the BH horizon
and the radial coordinate on the disk, and (\ref{eq3}) the
calculations of the poloidal and toroidal components of the magnetic
field at the disk. The criterion of SIBZ can be derived in an
analogous way. However, the BZ process involves unknown
astrophysical loads, to which both the mapping relation and the
calculations for the poloidal and toroidal components of the
magnetic field are related. In order to work out an analytical model
we present some simplified assumptions as follows.

(\ref{eq1}) The magnetosphere anchored in a Kerr BH and its
surrounding disk is described in Boyer-Lindquist coordinates, in
which the following Kerr metric parameters are involved (MacDonald
and Thorne 1982, hereafter MT82).

\begin{equation}
\label{eq1} \left\{ {\begin{array}{l}
 \Sigma ^2 = \left( {r^2 + a^2} \right)^2 - a^2\Delta \sin ^2\theta ,\mbox{
}\rho ^2 = r^2 + a^2\cos ^2\theta ,\mbox{ } \\
 \Delta = r^2 + a^2 - 2Mr,\mbox{ }\varpi = \left( {\Sigma \mathord{\left/
{\vphantom {\Sigma \rho }} \right. \kern-\nulldelimiterspace} \rho }
\right)\sin \theta , \\
 \alpha = {\rho \sqrt \Delta } \mathord{\left/ {\vphantom {{\rho \sqrt
\Delta } \Sigma }} \right. \kern-\nulldelimiterspace} \Sigma . \\
 \end{array}} \right.
\end{equation}

(\ref{eq2}) The remote load is axisymmetric, being located evenly in
a plane with some height above the disk. In Figure 1 the open
magnetic field lines connect the BH horizon with the load. The
symbol $L_{BZ} $ and $H_c $ represent the critical field line and
the height of the remote load above the equatorial plane for the
occurrence of SIBZ, respectively.

(\ref{eq3}) In Figure 1 the radius $r _{_S} $ is the critical radius
of SIMC, which is determined by the criterion of the screw
instability given in W04,

\begin{equation}
\label{eq2} {\left( {{2\pi \varpi _{_D} } \mathord{\left/ {\vphantom
{{2\pi \varpi _{_D} } {L_{MC} }}} \right. \kern-\nulldelimiterspace}
{L_{MC} }} \right)B_D^p } \mathord{\left/ {\vphantom {{\left( {{2\pi
\varpi _{_D} } \mathord{\left/ {\vphantom {{2\pi \varpi _{_D} }
{L_{MC} }}} \right. \kern-\nulldelimiterspace} {L_{MC} }}
\right)B_D^p } {B_D^T }}} \right. \kern-\nulldelimiterspace} {B_D^T
} < 1.
\end{equation}

\noindent In equation (\ref{eq2}) $L_{MC} $ is the critical length
of the poloidal field line for SIMC, and $B_D^p $ and $B_D^T $ are
the poloidal and toroidal components of the magnetic field on the
disk, respectively, and $\varpi _{_D} $ is the cylindrical radius on
the disk and it reads


\begin{equation}
\label{eq3} \varpi _{_D} = \Sigma_{D} / \rho _{_D}  = \xi M\chi
_{ms}^2 \sqrt {1 + a_ * ^2 \xi ^{ - 2}\chi _{ms}^{ - 4} + 2a_ * ^2
\xi ^{ - 3}\chi _{ms}^{ - 6} } .
\end{equation}

\noindent where $\chi _{ms} \equiv \sqrt {{r_{ms} } \mathord{\left/
{\vphantom {{r_{ms} } M}} \right. \kern-\nulldelimiterspace} M} $ is
defined by Novikov {\&} Thorne (1973) in terms of the radius of
innermost stable circular orbit (ISCO).

(\ref{eq4}) The angle $\theta _S $ in Figure 1 is the half-opening
angle of the magnetic flux tube on the horizon, which is related by
the mapping relation between the angular coordinate on the BH
horizon and the radial coordinate on the disk as follows (Wang et
al., 2003, hereafter W03),

\begin{equation}
\label{eq4} \cos \theta - \cos \theta _L = \int_1^\xi
{\mbox{G}\left( {a_ * ;\xi ,n} \right)d\xi } ,
\end{equation}

\noindent where

\begin{equation}
\label{eq5} \mbox{G}\left( {a_ * ;\xi ,n} \right) = \frac{\xi ^{1 -
n}\chi _{ms}^2 \sqrt {1 + a_ * ^2 \chi _{ms}^{ - 4} \xi ^{ - 2} +
2a_ * ^2 \chi _{ms}^{ - 6} \xi ^{ - 3}} }{2\sqrt {\left( {1 + a_ *
^2 \chi _{ms}^{ - 4} + 2a_ * ^2 \chi _{ms}^{ - 6} } \right)\left( {1
- 2\chi _{ms}^{ - 2} \xi ^{ - 1} + a_ * ^2 \chi _{ms}^{ - 4} \xi ^{
- 2}} \right)} }.
\end{equation}

\noindent In equations (\ref{eq4}) and (\ref{eq5}) $\xi \equiv r
\mathord{\left/ {\vphantom {r {r_{ms} }}} \right.
\kern-\nulldelimiterspace} {r_{ms} }$ is defined as a radial
parameter in terms of $r_{ms} $, and $n$ is a power-law index for
the variation of the poloidal magnetic field at the disk, i.e.,

\begin{equation}
\label{eq6} B_D^p \propto \xi ^{ - n}.
\end{equation}

(\ref{eq5}) The suspended accretion state is assumed due to the
transfer of angular momentum from the BH to the disk (van Putten
{\&} Ostriker, 2001).

Analogous to equation (\ref{eq2}) the criterion for SIBZ can be
expressed as

\begin{equation}
\label{eq7} {\left( {{2\pi R} \mathord{\left/ {\vphantom {{2\pi R}
{L_{BZ} }}} \right. \kern-\nulldelimiterspace} {L_{BZ} }}
\right)B_L^p } \mathord{\left/ {\vphantom {{\left( {{2\pi R}
\mathord{\left/ {\vphantom {{2\pi R} {L_{BZ} }}} \right.
\kern-\nulldelimiterspace} {L_{BZ} }} \right)B_L^p } {B_L^T }}}
\right. \kern-\nulldelimiterspace} {B_L^T } < 1,
\end{equation}

\noindent where $L_{BZ} $ is the critical length of the poloidal
field line for SIBZ, and $B_L^p $ and $B_L^T $ are the poloidal and
toroidal components of the magnetic field on the remote load,
respectively, and $R$ is the cylindrical radius of the remote load
with respect to the symmetric axis of the BH.

The toroidal field component $B_L^T $ can be expressed by Ampere's
law,

\begin{equation}
\label{eq8} B_L^T = {2I_L } \mathord{\left/ {\vphantom {{2I_L } R}}
\right. \kern-\nulldelimiterspace} R,
\end{equation}

\noindent where $I_L $ is the electric current flowing in the loop
$KM{M}'{K}'$ in Figure 1 and it reads

\begin{equation}
\label{eq9} I_L = \sqrt {{P_{BZ} } \mathord{\left/ {\vphantom
{{P_{BZ} } {Z_L }}} \right. \kern-\nulldelimiterspace} {Z_L }} .
\end{equation}

The quantities $P_{BZ} $ and $Z_L $ in equation (\ref{eq9}) are the
BZ power and the load resistance, respectively. The BZ power has
been derived in W02 as follows,

\begin{equation}
\label{eq10} {P_{BZ} } \mathord{\left/ {\vphantom {{P_{BZ} } {P_0
}}} \right. \kern-\nulldelimiterspace} {P_0 } = 2a_ * ^2
\int_0^{\theta _S } {\frac{k\left( {1 - k} \right)\sin ^3\theta
d\theta }{2 - \left( {1 - q} \right)\sin ^2\theta }} ,
\end{equation}

\noindent where $q \equiv \sqrt {1 - a_ * ^2 } $ is a parameter
depending on the BH spin, and $k \equiv {\Omega _F } \mathord{\left/
{\vphantom {{\Omega _F } {\Omega _{_H} }}} \right.
\kern-\nulldelimiterspace} {\Omega _{_H} }$ is the ratio of the
angular velocity of the field lines to that of the BH horizon. The
quantity $P_0 $ is defined by

\begin{equation}
\label{eq11} P_0 \equiv \left( {B_H^p } \right)^2M^2 \approx
6.59\times 10^{50}\times B_{15}^2 \left( {M \mathord{\left/
{\vphantom {M {M_ \odot }}} \right. \kern-\nulldelimiterspace} {M_
\odot }} \right)^2erg \cdot s^{ - 1},
\end{equation}

\noindent where $B_{15} $ represents the magnetic field at the BH
horizon in terms of $10^{15}gauss$.

MT82 argued in a speculative way that the ratio $k$ will be
regulated to about 0.5 by the BZ process itself, which corresponds
to the optimal BZ power with the impedance matching. Taking the
impedance matching into account, we have the remote load resistance
equal to the horizon resistance, and they read

\begin{equation}
\label{eq12} \Delta Z_L = \Delta Z_H = R_H \frac{\rho _{_H} d\theta
}{2\pi \varpi _{_H} },
\end{equation}

\noindent where $R_H = 4\pi = 377ohm$ is the surface resistivity of
the BH horizon (MT82). Thus we have $Z_L $ and $Z_H $ expressed as

\begin{equation}
\label{eq13} Z_L = Z_H = \int_0^{\theta _S } {R_H \frac{\rho _{_H}
d\theta }{2\pi \varpi _{_H} }} = \int_0^{\theta _S } {\frac{2\rho
_{_H} d\theta }{\varpi _{_H} }}
\end{equation}

Incorporating equations (\ref{eq8})---(\ref{eq13}), we can calculate
$B_L^T $ in terms of the cylindrical radius $R$. On the other hand,
the poloidal magnetic field $B_L^P $ at the radius $R$ of the load
can be determined by the conservation of the magnetic flux, i.e.,

\begin{equation}
\label{eq14} B_H^P 2\pi \varpi _{_H} \rho _{_H} d\theta = B_L^P 2\pi
RdR.
\end{equation}

\noindent From equation (\ref{eq14}) we have

\begin{equation}
\label{eq15} B_L^P = \frac{B_H^P \varpi _{_H} \rho _{_H}
}{R}\frac{d\theta }{dR}.
\end{equation}

Assuming that the height of the planar load above the equatorial
plane of the Kerr BH is$ H$, we have an approximate relation between
the angle $\theta $ and the radius $R$ as follows,

\begin{equation}
\label{eq16} \tan \theta = R \mathord{\left/ {\vphantom {R H}}
\right. \kern-\nulldelimiterspace} H.
\end{equation}

\noindent Substituting equation (\ref{eq16}) into equation
(\ref{eq15}), we have

\begin{equation}
\label{eq17} B_L^P = \frac{B_H^P \varpi _{_H} \rho _{_H} \cos
^2\theta }{HR}.
\end{equation}

\noindent Incorporating equations (\ref{eq8}) and (\ref{eq17}) with
the criterion (\ref{eq7}), we have

\begin{equation}
\label{eq18} \frac{\pi B_H^P \varpi _{_H} \rho _{_H} \sin \theta _S
\cos ^2\theta _S }{H\sqrt {{P_{BZ} } \mathord{\left/ {\vphantom
{{P_{BZ} } {Z_L }}} \right. \kern-\nulldelimiterspace} {Z_L }} } <
1,
\end{equation}

\noindent where the relation $\sin \theta _S = R \mathord{\left/
{\vphantom {R {L_{BZ} }}} \right. \kern-\nulldelimiterspace} {L_{BZ}
}$ is used. The criterion (\ref{eq18}) implies that SIBZ will occur,
provided that the height of the load is greater than the critical
height, i.e., $H > H_c $, and we have

\begin{equation}
\label{eq19} h_c \equiv {H_c } \mathord{\left/ {\vphantom {{H_c }
M}} \right. \kern-\nulldelimiterspace} M = \frac{\pi B_H^P \varpi
_{_H} \rho _{_H} \sin \theta _S \cos ^2\theta _S }{M\sqrt {{P_{BZ} }
\mathord{\left/ {\vphantom {{P_{BZ} } {Z_L }}} \right.
\kern-\nulldelimiterspace} {Z_L }} }.
\end{equation}

As argued in W04 the angle $\theta _S $ can be determined by the
criterion of SIMC with the mapping relation between the BH horizon
and the disk, and it is a function of the BH spin $a_ * $ and the
power-law index $n$, i.e., $\theta _S \left( {a_ * ,n} \right)$.
Inspecting equations (\ref{eq1}), (\ref{eq10}) and (\ref{eq19}), we
find that $h_c $ is a dimensionless parameter also depending on the
parameters $a_ * $ and $n$, i.e., $h_c = h_c \left( {a_ * ,n}
\right)$. By using equations (\ref{eq2}) and (\ref{eq19}) we have
the contours of $\theta _S \left( {a_ * ,n} \right)$ and $h_c \left(
{a_ * ,n} \right)$ in $a_ * - n$ parameter space as shown in Figure
2.

Inspecting Figure 2, we find the following features of the contours:

(\ref{eq1}) The values of $\theta _S $ increases and those of $h_c $
decreases with the increasing $n$ for the given BH spin $a_ * $,
respectively.

(\ref{eq2}) The values of $\theta _S $ increases and those of $h_c $
decreases with the increasing $a_ * $ for the given $n$,
respectively.

(\ref{eq3}) Both SIMC and SIBZ will occur, provided that the
parameters $a_ * $, $n$ and $h_c $ are greater than some critical
values. As shown in Figure 2c the shaded region indicates the value
ranges of $a_ * $ and $n$ in which both $\theta _S > 0$ and $h_c >
100$ are constrained. Thus the occurrence of SIMC and SIBZ is
guaranteed by the value ranges of $a_ * $ and $n$ in the shaded
region.

\section{TIME SCALE OF A GRB AND ENERGY EXTRACTION FROM A ROTATING BH}

There are several scenarios to invoke the BZ process for powering
GRBs, and the main differences among these scenarios lie in the
environment of a spinning BH and the approaches to the duration of a
GRB. These scenarios are outlined as follows.

\textbf{Model I}: In Lee00 the energy is extracted magnetically from
a rotating BH without disk, and the duration of a GRB is estimated
as the time for extracting all rotational energy of the central BH
via the BZ process.

\textbf{Model II}: It is argued that the energy is extracted
magnetically from a rotating BH with a transient disk, and the
duration of a GRB is estimated as the time for the disk plunged into
the BH (Lee {\&} Kim 2002; Wang et al. 2002).

\textbf{Model III}: In Li00 the energy is extracted magnetically
from a rotating BH with a stationary torus in the state of suspended
accretion, and the duration of a GRB is estimated roughly as the
time for extracting all rotational energy of the central BH via the
BZ process.

\textbf{Model IV}: In B00 the energy is extracted magnetically from
a rotating BH with a non-stationary disk in the state of suspended
accretion, and the duration of a GRB is determined by the presence
of the disk.

\textbf{Model V}: In P03 the energy is extracted magnetically from a
rotating BH with a torus in the state of suspended accretion, and
the duration of a GRB is determined by the instability of the disk.

\textbf{Model VI}: In Lei05 the energy is extracted magnetically
from a rotating BH with a thin disk in the state of suspended
accretion, and the duration of a GRB is determined by the lifetime
of the half-opening angle.

\textbf{This Model:} It is a modified version of Model VI, in which
the effects of SIMC and SIBZ are taken into account. Compared with
Model VI some features and advantages are given as follows.

(\ref{eq1}) The magnetic field configuration in Model VI is built
based on the conservation of the closed magnetic flux connecting the
BH with the disk with the precedence over the open magnetic flux,
resulting in the closed field lines connecting the half-open angle
$\theta _{BZ} $ at horizon with the disk extending to infinity. In
this Model, however, the closed field lines are confined by SIMC
within a region of a limited radius $r _{_S} $ (about a few
Schwarzschild radii as shown in Table 4), which is consistent with
the collapsar model for GRBs-SNe.

(\ref{eq2}) As argued in the next section, the variability time
scales of the light curves of GRBs are modulated by two successive
flares due to SIBZ.

The main features of the above models for GRBs are summarized in
Table 1.

In Model VI the duration of a GRB is regarded as the lifetime of the
half-opening angle $\theta _{BZ} $, which is based on the evolution
of the rotating BH. The same procedure can be applied to this model
except that the angle $\theta _{BZ} $ is replaced by $\theta _S $
arising from SIMC. It is found that the characteristics of the BH
evolution in this model are almost the same as given in Model VI as
shown by $a_ * - n$ parameter spaces in Figure 3.

For several GRB-SNe, the observed energy $E_\gamma $ and duration
$T_{90} $ can be fitted by adjusting the parameters $n$ and $B_{15}
$. The energy $E_{SN} $ can be predicted as shown in Table 2, where
the values of $n$, $B_{15} $ and $E_{SN} $ fitted to five GRBs
invoking Model VI and this model are shown in the left and right
sub-columns, respectively.

The fractions of extracting energy from a rotating BH via the BZ and
MC processes are defined respectively as $f_{BZ} $ and $f_{MC} $,
and they read

\begin{equation}
\label{eq20} f_{BZ} = \frac{E_{BZ} }{E_{BZ} + E_{MC} }, \quad f_{MC}
= \frac{E_{MC} }{E_{BZ} + E_{MC} },
\end{equation}

\noindent where $E_{BZ} $ and $E_{MC} $ are the energies extracting
in the BZ and MC processes, respectively.

\begin{equation}
\label{eq21} E_{BZ} = \int_0^{t_{BZ} } {P_{BZ} dt} , \quad E_{MC} =
\int_0^{t_{BZ} } {P_{MC} dt} .
\end{equation}

In equation (\ref{eq21}) $t_{BZ} $ is defined as the lifetime of the
angle $\theta _S $, which can be calculated by the same procedure
given in Lei05.\textbf{ }The MC power in equation (\ref{eq21}) is
expressed as (W04)

\begin{equation}
\label{eq22} {P_{MC} } \mathord{\left/ {\vphantom {{P_{MC} } {P_0
}}} \right. \kern-\nulldelimiterspace} {P_0 } = 2a_ * ^2
\int_{\theta _S }^{\theta _L } {\frac{\beta \left( {1 - \beta }
\right)\sin ^3\theta d\theta }{2 - \left( {1 - q} \right)\sin
^2\theta }} ,
\end{equation}

\noindent where the parameter $\beta \equiv {\Omega _F }
\mathord{\left/ {\vphantom {{\Omega _F } {\Omega _{_H} }}} \right.
\kern-\nulldelimiterspace} {\Omega _{_H} } = {\Omega _{_D} }
\mathord{\left/ {\vphantom {{\Omega _{_D} } {\Omega _{_H} }}}
\right. \kern-\nulldelimiterspace} {\Omega _{_H} }$ is the ratio of
the angular velocity of the magnetic field lines to that of the BH.

In Model VI the BH spin $a_\ast ^{GRB} $ corresponds to the time
when the half-opening angle $\theta _{BZ} = 0$, and the BZ power
$P_{BZ} = 0$, while $a_\ast ^{SN} $ corresponds to the MC power
$P_{MC} = 0$. In this model $a_\ast ^{GRB} $ and $a_\ast ^{SN} $
have the same meanings as given in Model VI except that $\theta
_{BZ} $ is replaced by $\theta _S $.

As shown in Table 3, the values of $a_\ast ^{GRB} $, $f_{BZ} $ and
$f_{MC} $ fitted to five GRBs invoking Model VI and this model are
shown in the left and right sub-columns, respectively.

Inspecting the data in the left and right sub-columns in Table 2 and
Table 3, we find that $f_{BZ} $ and $f_{MC} $ in this model are
greater and less than the counterparts in Model VI, respectively.
This result arises from the effects of SIMC and SIBZ: the
half-opening angle $\theta _S $ in this model is greater than the
half-opening angle $\theta _{BZ} $ in Model VI as argued in W04.

\section{AN EXPLANATION FOR VARIABILITIES IN GRB LIGHT CURVES}

As is well known, the bursts are divided into long and short bursts
according to their $T_{90}$. Most GRBs are highly variable, showing
100{\%} variations in flux on a time scale much shorter than the
overall duration of the burst. The bursts seem to be composed of
individual pulses, with a pulse being the ``building block'' of the
overall light curve. The variability time scale $\delta t $is much
shorter than the GRBs' duration $T_{90}$, the former is more than a
factor of 10$^{4}$ smaller than the latter (Piran, 2004). However,
the origin of the variability in the light curves of GRBs remains
unclear.

In this paper, we combine the variability with the screw instability
of the magnetic field in the BH magnetosphere, and suggest that the
variability could be fitted by a series of flares arising from SIBZ,
which accompanies the release of the energy of the toroidal magnetic
field.

An equivalent circuit $ML{L}'{M}'$ for SIBZ is shown in Figure 4a,
which consisting of two adjacent magnetic surfaces $M{M}'$ and
$L{L}'$ connecting the BH horizon and the remote load. An inductor
is introduced in the equivalent circuit by considering that the
toroidal magnetic field threads the loop $ML{L}'{M}'$, and the
inductor is represented by the symbol $\Delta L$ in Figure 4a.

The inductance $\Delta L$ in the circuit is defined by

\begin{equation}
\label{eq23} \Delta L = {\Delta \Psi ^T} \mathord{\left/ {\vphantom
{{\Delta \Psi ^T} {I_L }}} \right. \kern-\nulldelimiterspace} {I_L
},
\end{equation}

\noindent where $I_L $ is given by equation (\ref{eq9}), and $\Delta
\Psi ^T$ is the flux of the toroidal magnetic field threading the
circuit. The flux $\Delta \Psi ^T$ can be integrated over the loop
$ML{L}'{M}'$ as follows,

\begin{equation}
\label{eq24} \Delta \Psi ^T = \oint_{loop} {B^T\sqrt {g_{rr}
g_{\theta \theta } } dr} d\theta ,
\end{equation}

\noindent where the toroidal magnetic field measured by
``zero-angular-momentum observers'' is

\begin{equation}
\label{eq25} B^T = {2I_L } \mathord{\left/ {\vphantom {{2I_L }
{\left( {\alpha \varpi } \right)}}} \right.
\kern-\nulldelimiterspace} {\left( {\alpha \varpi } \right)},
\end{equation}

\noindent where $\alpha $ is the lapse function defined in equation
(\ref{eq1}) (MT82).

Since the geometric shapes of the magnetic surfaces are unknown, we
assume that the surfaces are formed by rotating the two radial
segments $M{M}'$ and $L{L}'$, which span the angle $\Delta \theta $
as shown in Figure 4b. Thus the flux $\Delta \Psi ^T$ can be
calculated easily by integrating over the region $ML{L}'{M}'$.
Incorporating equations (\ref{eq23})---(\ref{eq25}), we obtain
$\Delta L$as follows.

\begin{equation}
\label{eq26} \Delta L = 2\csc \theta \Delta \theta \int_{r_H
}^{L_{BZ} } {{\rho ^2dr} \mathord{\left/ {\vphantom {{\rho ^2dr}
\Delta }} \right. \kern-\nulldelimiterspace} \Delta } = 2M\Delta
\theta \csc \theta _S \int_{\left( {1 + q} \right)}^{{L_{BZ} }
\mathord{\left/ {\vphantom {{L_{BZ} } M}} \right.
\kern-\nulldelimiterspace} M} {\frac{\left( {\tilde {r}^2 + a_
* ^2 \cos ^2\theta _S } \right)}{\left( {\tilde {r}^2 + a_ * ^2 - 2\tilde
{r}} \right)}d\tilde {r}} ,
\end{equation}

\noindent where $\tilde {r}$ is defined as $\tilde {r} \equiv r
\mathord{\left/ {\vphantom {r M}} \right. \kern-\nulldelimiterspace}
M$.

Although the detailed process of SIBZ is still unclear, we suggest
that the energy release in one event of SIBZ is roughly divided into
two stages: the processes for releasing and retrieving magnetic
energy, respectively. The two processes can be simulated as the
corresponding processes in the equivalent$ R-L $circuit. The
detailed analysis is given as follows.

At the first stage the energy of the toroidal magnetic field is
released as soon as SIBZ occurs, being dissipated on the load and
plasma fluid in the way analogous to a discharging process in an
equivalent $R-L$ circuit. The equation governing the discharging
process in $R-L$ circuit is

\begin{equation}
\label{eq27} \Delta L\frac{dI^P}{dt} + \left( {\Delta Z_{PLSM} +
\Delta Z_L } \right)I^P = 0,
\end{equation}

\noindent where $\Delta Z_{PLSM} $ is the resistance of the plasma
fluid in the BH magnetosphere.

At the second stage the energy of toroidal magnetic field is
recovered due to the rotation of the BH, and the process for
retrieving magnetic energy is modulated by a charging process in an
equivalent $R-L$ circuit. The equation governing the charging
process in $R-L$ circuit is

\begin{equation}
\label{eq28} \Delta L\frac{dI^P}{dt} + \left( {\Delta Z_H + \Delta
Z_L } \right)I^p = \Delta \varepsilon _{_H} .
\end{equation}

In equations (\ref{eq27}) and (\ref{eq28}) $\Delta L$ is the
inductance in the circuit $ML{L}'{M}'$, and $\Delta Z_{PLSM} $ is
the resistance of the plasma in the BH magnetosphere. Incorporating
equations (\ref{eq12}) and (\ref{eq26}), we have

\begin{equation}
\label{eq29} {\Delta L} \mathord{\left/ {\vphantom {{\Delta L}
{\Delta Z_H }}} \right. \kern-\nulldelimiterspace} {\Delta Z_H } =
\left( {9.85\times 10^{ - 6}\sec } \right)\frac{\left( {M
\mathord{\left/ {\vphantom {M {M_ \odot }}} \right.
\kern-\nulldelimiterspace} {M_ \odot }} \right)}{2 - \left( {1 - q}
\right)\sin ^2\theta _S }\int_{\left( {1 + q} \right)}^{{L_{BZ} }
\mathord{\left/ {\vphantom {{L_{BZ} } M}} \right.
\kern-\nulldelimiterspace} M} {\frac{\left( {\tilde {r}^2 + a_ * ^2
\cos ^2\theta _S } \right)}{\left( {\tilde {r}^2 + a_ * ^2 - 2\tilde
{r}} \right)}d\tilde {r}} .
\end{equation}

Combining the initial conditions in the first and second stages, we
have the solutions of equations (\ref{eq27}) and (\ref{eq28}) as
follows,

\begin{equation}
\label{eq30} I_{disch}^p = I_{initial}^p e^{ - t \mathord{\left/
{\vphantom {t {\tau _1 }}} \right. \kern-\nulldelimiterspace} {\tau
_1 }},
\end{equation}

\begin{equation}
\label{eq31} I_{ch}^p = I_{steady}^p \left( {1 - e^{ - t
\mathord{\left/ {\vphantom {t {\tau _2 }}} \right.
\kern-\nulldelimiterspace} {\tau _2 }}} \right).
\end{equation}

In equations (\ref{eq30}) and (\ref{eq31}) $I_{initial}^p $ and
$I_{steady}^p $ are the initial and steady currents, respectively,
while $I_{disch}^p $ and $I_{ch}^p $ represent the discharging and
charging currents, respectively. The characteristic time scales in
equations (\ref{eq30}) and (\ref{eq31}) are given respectively by
$\tau _1 $ and $\tau _2 $, and they read

\begin{equation}
\label{eq32} \tau _1 \equiv {\Delta L} \mathord{\left/ {\vphantom
{{\Delta L} {\left( {\Delta Z_{PLSM} + \Delta Z_L } \right)}}}
\right. \kern-\nulldelimiterspace} {\left( {\Delta Z_{PLSM} + \Delta
Z_L } \right)},
\end{equation}

\begin{equation}
\label{eq33} \tau _2 \equiv {\Delta L} \mathord{\left/ {\vphantom
{{\Delta L} {\left( {\Delta Z_H + \Delta Z_L } \right)}}} \right.
\kern-\nulldelimiterspace} {\left( {\Delta Z_H + \Delta Z_L }
\right)}.
\end{equation}

\noindent From equations (\ref{eq32}) and (\ref{eq33}) we have the
ratio of $\tau _1 $ to $\tau _2 $ given by

\begin{equation}
\label{eq34} {\tau _1 } \mathord{\left/ {\vphantom {{\tau _1 } {\tau
_2 }}} \right. \kern-\nulldelimiterspace} {\tau _2 } = {2\Delta Z_H
} \mathord{\left/ {\vphantom {{2\Delta Z_H } {\left( {\Delta
Z_{PLSM} + \Delta Z_H } \right)}}} \right.
\kern-\nulldelimiterspace} {\left( {\Delta Z_{PLSM} + \Delta Z_H }
\right)},
\end{equation}

\noindent where $\Delta Z_H = \Delta Z_L $ is used in deriving
equation (\ref{eq34}).

In contrast to disk plasma of perfect conductivity, the resistance
$\Delta Z_{PLSM} $ cannot be neglected based on the following
considerations:

(\ref{eq1}) The plasma fluid becomes very tenuous after leaving the
inner edge of the disk, augmenting significantly the resistance due
to an increasing radial velocity onto the BH;

(\ref{eq2}) The conductivity of the plasma fluid is highly
anisotropic, i.e., the conductivity in the cross-field direction is
greatly impeded by the presence of the strong magnetic threading the
BH (Punsly 2001).

Although the value of $\Delta Z_{PLSM} $ is unknown, we can estimate
roughly the variability time scales of GRBs by combining equation
(\ref{eq34}) with different cases given as follows.

\textbf{CASE I: }$\Delta Z_{PLSM} > > \Delta Z_H $ leads to $\tau _1
< < \tau _2 $, and the time scale of two successive flares arising
from SIBZ is dominated by $\tau _2 $.

\textbf{CASE II: }$\Delta Z_{PLSM} \approx \Delta Z_H $ leads to
$\tau _1 \approx \tau _2 $, and the time scale of two successive
flares arising from SIBZ is about $2\tau _2 $.

\textbf{CASE III: }$\Delta Z_{PLSM} < < \Delta Z_H $ leads to $\tau
_1 \approx 2\tau _2 $, and the time scale of two successive flares
arising from SIBZ is about $3\tau _2 $.

Therefore the variability time scales are insensitive to the values
of the unknown $\Delta Z_{PLSM} $. From equation (\ref{eq31}) we
obtain that the charging current attains 99.3{\%} of $I_{steady}^p $
in the relax time $t_{relax} = 5\tau _2 $, implying the recovery of
the toroidal magnetic field, and the variability time scales of GRBs
can be estimated as follows,

\begin{equation}
\label{eq35} \left( {\delta t} \right)_I \equiv \left( {t_{SIBZ} }
\right)_I \approx 5\tau _2 ,\quad for \quad \Delta Z_{PLSM} > >
\Delta Z_H ,
\end{equation}

\begin{equation}
\label{eq36} \left( {\delta t} \right)_{II} \equiv \left( {t_{SIBZ}
} \right)_{II} \approx 10\tau _2 ,\quad for \quad \Delta Z_{PLSM}
\approx \Delta Z_H ,
\end{equation}

\begin{equation}
\label{eq37} \left( {\delta t} \right)_{III} \equiv \left( {t_{SIBZ}
} \right)_{III} \approx 15\tau _2 ,\quad for \quad \Delta Z_{PLSM} <
< \Delta Z_H . \end{equation}

In \textbf{CASE I} we have $\tau _1 < < \tau _2 $, implying that the
magnetic energy is released much more rapidly in the first stage
compared with the time for the recovery of the magnetic energy in
the second stage. Thus \textbf{CASE I} seems more consistent with
the feature of the light curves, i.e., an individual pulse is a
fast-rise exponential decay (FRED) with an average rise-to-decay
ratio of 1:3 (Norris et al. 1996).

By using equations (\ref{eq29}), (\ref{eq33}) and (35)---(37) we
have the variability time scales in the light curves of four GRBs in
the three different cases as shown in Table 4. In addition, we
obtain the curves of $\left( {\delta t} \right)_I $ versus $a_ * $
with the fixed values of $n$ for GRB 990712, GRB 991208 and GRB
021216 as shown in Figure 5.

Inspecting Table 4, we find that the variability time scales of tens
of msec in the light curves of GRBs can be modulated by the two
successive flares due to SIBZ, which accompany the BZ process in
powering the GRBs. We also find that the variability time scales of
the four GRBs are generally three order of magnitude less than the
corresponding durations $T_{90} $, which are consistent with the
observations.

From Figure 5 we find that the curves of $\left( {\delta t}
\right)_I $ versus $a_ * $ are almost the same, increasing linearly
in a rather small slope with the decreasing $a_ * $. The values of
$\left( {\delta t} \right)_I $ remains less than 20 msec during the
occurrence of SIBZ for these GRBs.

\section{DISCUSSION}

\subsection{Mechanism for the recurrent occurrence of SIBZ}

In this paper we discuss the possibility of the modulations of SIBZ
on the light curves of GRBs. One of the puzzles is what mechanism
leads to the recurrent occurrence of SIBZ, and prevents the magnetic
field from settling to a screw-stable configuration.

As argued in MT82 the BH magnetosphere consists of a series of
magnetic surfaces connecting the horizon with the loads, and the
total electric current $I$ flowing downward through an
\textbf{\textit{m}}-loop is proportional to the toroidal magnetic
field $B^T$ by Ampere's law. It is argued that these magnetic
surfaces can be regarded as an equivalent circuit, in which each
loop consists of two adjacent magnetic surfaces (W02). The total
electric current flowing downward through an
\textbf{\textit{m}}-loop is exactly equal to the algebraic sum of
the poloidal currents flowing in the loops (W03).

As shown in Figure 1 the critical magnetic surface (henceforth CMS)
for SIBZ is represented by the critical line $M{M}'$. According to
the criterion (\ref{eq7}) only the toroidal magnetic field outside
CMS is depressed by SIBZ, while the toroidal and poloidal components
of the magnetic field within CMS are little affected. On the other
hand, the poloidal magnetic field outside CMS still exists in spite
of the occurrence of SIBZ, and the toroidal magnetic field outside
CMS will be recovered because of the twist of the poloidal magnetic
field arising from the rotation of the BH. So, the rotation of the
BH is the main mechanism for the recurrent occurrence of SIBZ in the
BH magnetosphere.

\subsection{Rotation period of BH and time scales of recovery of
toroidal magnetic fields}

If we take BH mass as 10$M_ \odot $, the rotation period of the BH
is only $\sim $1msec for the BH spin required by the criterion of
SIBZ. This result implies that toroidal magnetic fields can not be
recovered in one period of a rotating BH. How to explain the
discrepancy between BH rotation and the time scales required for
recovery of toroidal fields ?

In spite of lack of detailed knowledge of the screw instability, it
is helpful to imagine the magnetic field line as an elastic string.
The rotating BH always twists the field line, while the field line
tries to untwist itself. Once the toroidal component of the magnetic
field is strong enough to satisfy the criterion, the screw
instability will occurs, just as a twisted elastic string releases
its energy under appropriate conditions. In our model we simulate
the process for twisting the field line by a transient process for
accumulating magnetic energy in the inductor $\Delta L$ in the
equivalent $R-L$ circuit, which corresponds to the increase of the
toroidal magnetic field, and the variability time scales of tens of
msec in the light curves of several GRBs are fitted as the time
interval between two successive flares due to SIBZ. Thus the
threshold of toroidal magnetic field (magnetic energy) cannot be
recovered in only one rotation of BH, just as a threshold of
twisting more than one turn is required for an elastic string to
release its energy.

\
\subsection{An explanation for GRBs with XRFs and XRRs }

Recently, much attention has been paid to the issue of X-ray flashes
(XRFs), X-ray-rich gamma-ray bursts (XRRs) and GRBs, since HETE-2
provided strong evidence that the properties of these three kinds of
bursts form a continuum, and therefore these objects are probably
the same phenomenon (Lamb et al. 2004a, 2004b, 2005). The
observations from HETE-2 motivate some authors to seek a unified
model of these bursts. The most competitive unified models of these
bursts are off-axis jet model (Yamazaki, Ioka, {\&} Nakamura 2002;
Lamb et al. 2005) and two-component jet model (Berger et al. 2003;
Huang et al. 2004), in which XRFs, XRRs and GRBs arise from the
differences in the viewing angles. Unfortunately, a detailed
discussion on producing different viewing angles for XRFs, XRRs and
GRBs has not been given in the above works.

Our argument on SIBZ and SIMC may be helpful to understanding this
issue. It is believed that a disk is probably surrounded by a
high-temperature corona analogous to the solar corona (Liang {\&}
Price 1977; Haardt 1991; Zhang et al 2000). Very recently, some
authors argued that the coronal heating in some stars including the
Sun is probably related to dissipation of currents, and very strong
X-ray emissions arise from variation of magnetic fields (Galsgaard
{\&} Parnell 2004; Peter et al. 2004). Analogously, if the corona
exists above the disk in our model, we expect that it might be
heated by the induced currents due to SIMC and SIBZ. Therefore a
very strong X-ray emission would be produced to form XRFs or XRRs.

Although our model may be too simplified and idealized with respect
to the real situation, it provides a possible scenario for the
occurrence of the screw instability in BH magnetosphere, and it may
be helpful to understanding some astrophysical observations. We hope
to improve our model by combining more observations in the future.

\acknowledgments
 {\bf Acknowledgements:}
 This work is supported by the National Natural
Science Foundation of China under Grant Numbers 10373006, 10573006
and 10121503. The anonymous referee is thanked for his (her) helpful
comments and suggestions.


\begin{figure}

\epsscale{0.5}
\begin{center}
\plotone{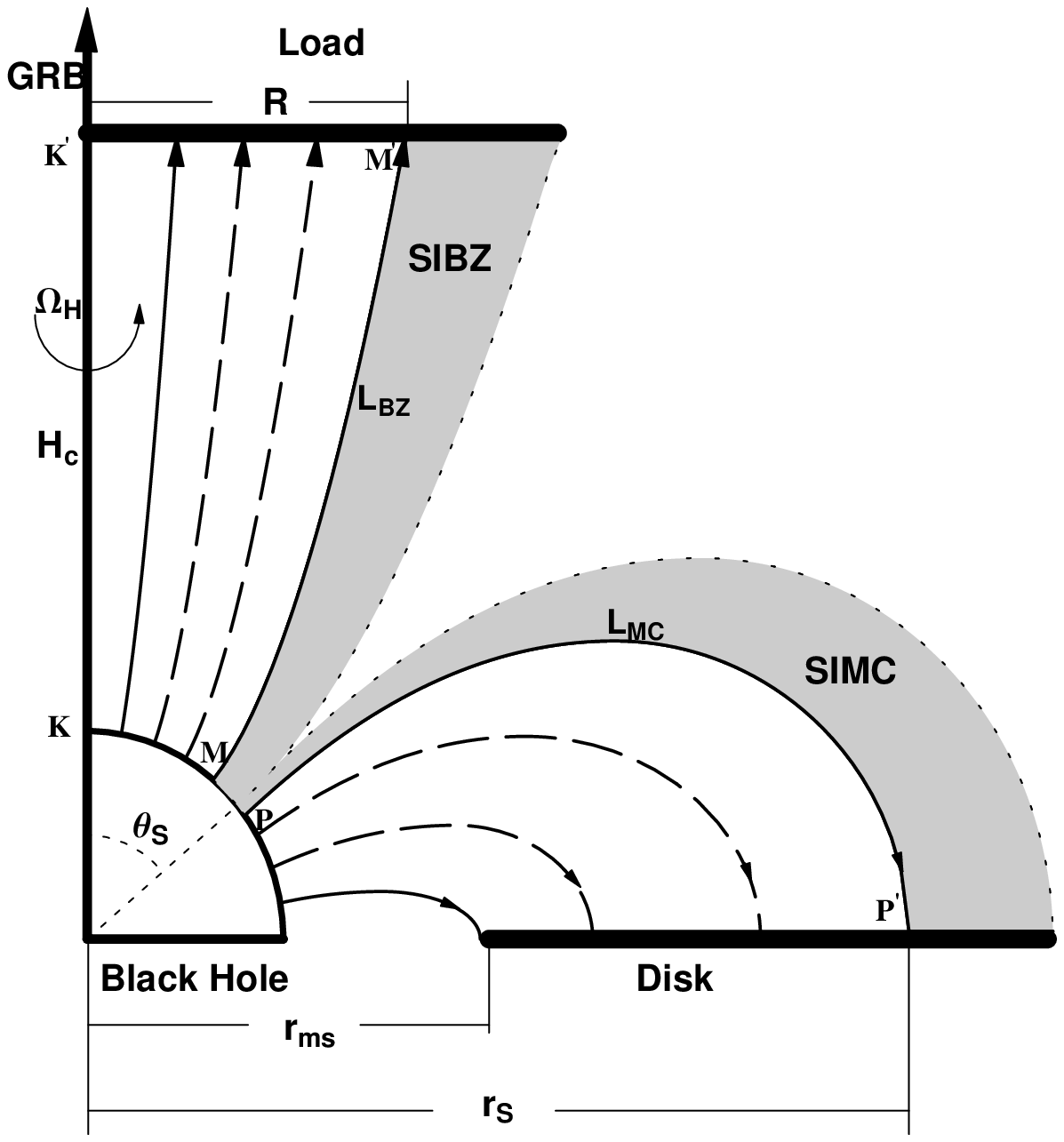}
\end{center}

\caption{A schematic drawing for magnetic field configuration with
screw instability in BH magnetosphere, where the screw-unstable
regions for SIBZ and SIMC are indicated by the shaded regions (not
in scale).} \label{fig1}

\end{figure}

\begin{center}
\begin{figure}
\epsscale{0.35}
\begin{center}
 \plotone{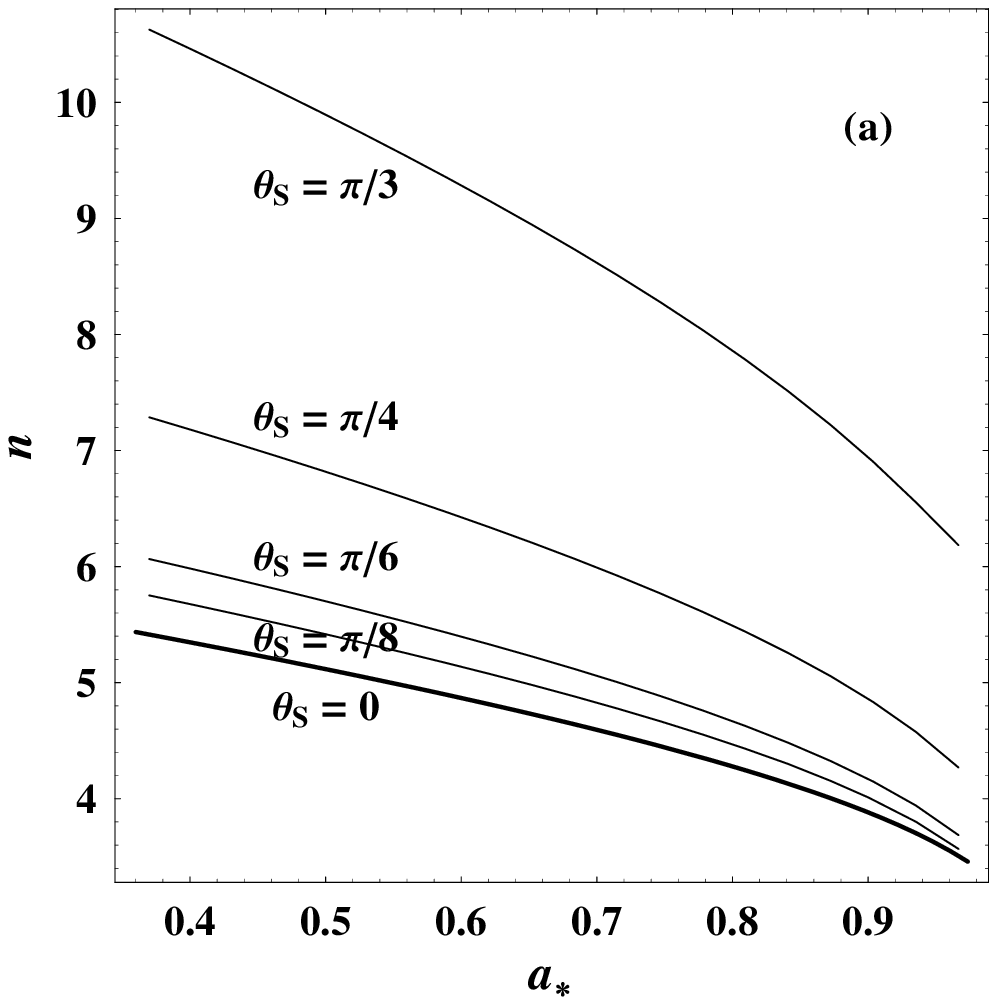}
\end{center}
\begin{center}
 \epsscale{0.35}
  \plotone{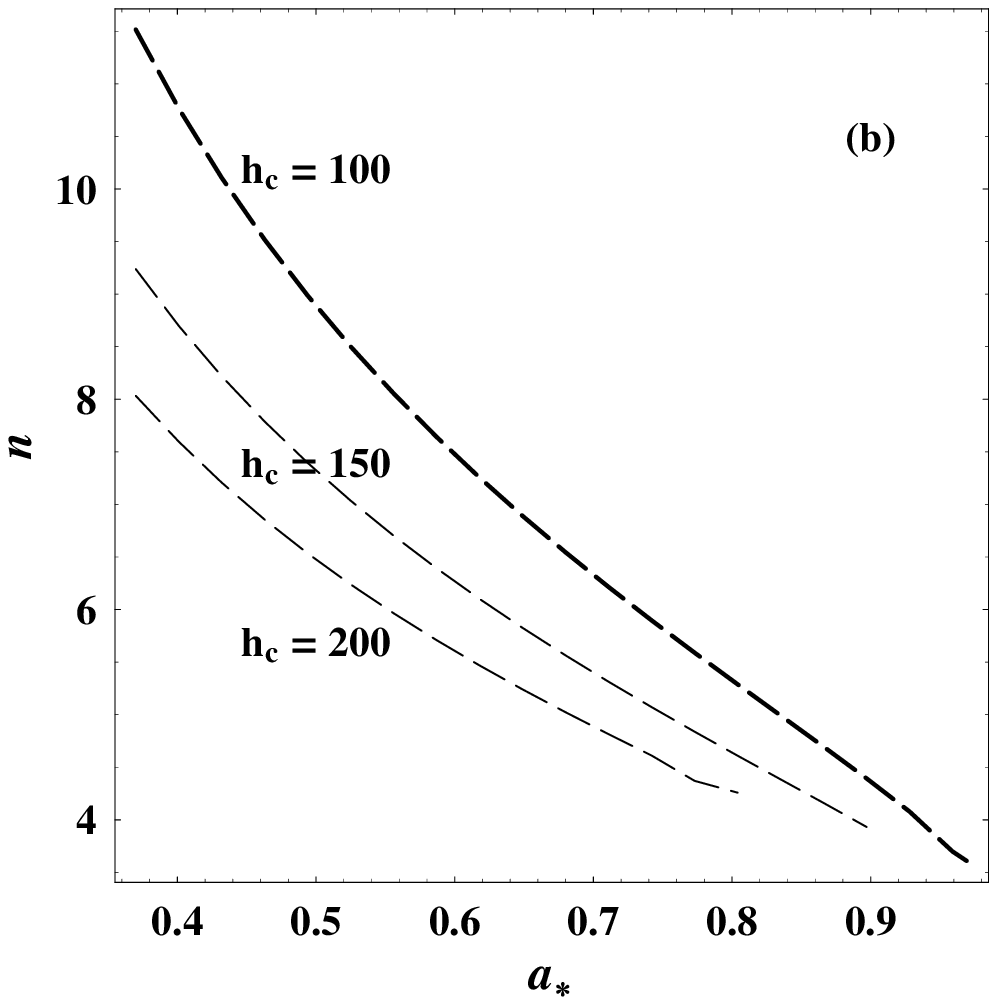}
\end{center}
\begin{center}
 \epsscale{0.35}
 \plotone{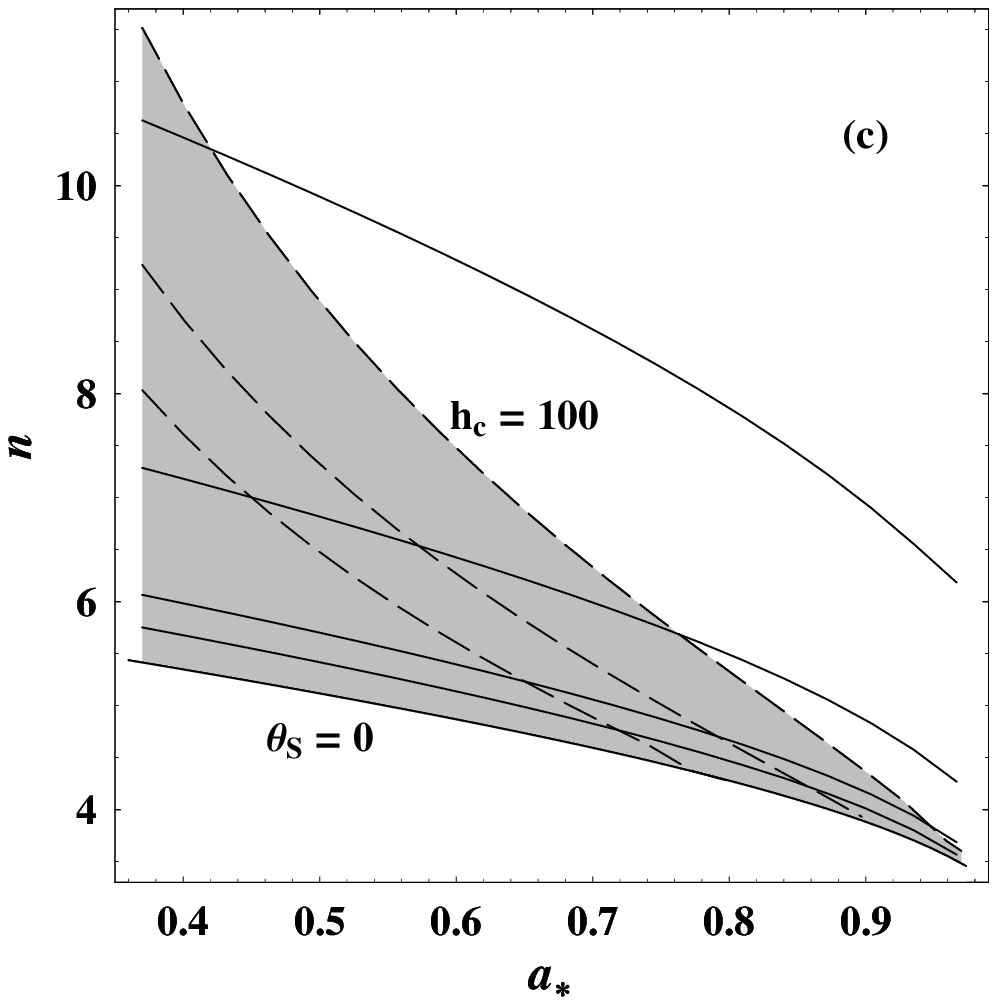}
 \end{center}
\caption{(a) The contours of $\theta _S \left( {a_ * ,n} \right)$
(solid lines); (b) the contours of $h_c \left( {a_
* ,n} \right)$ (dashed lines); (c) a shaded region for
occurrence of SIMC and SIBZ bounded by the contours $\theta _S = 0$
and $h_c = 100$.} \label{fig2}
\end{figure}
\end{center}

\begin{figure}
\begin{center}
\epsscale{0.4}
 \plotone{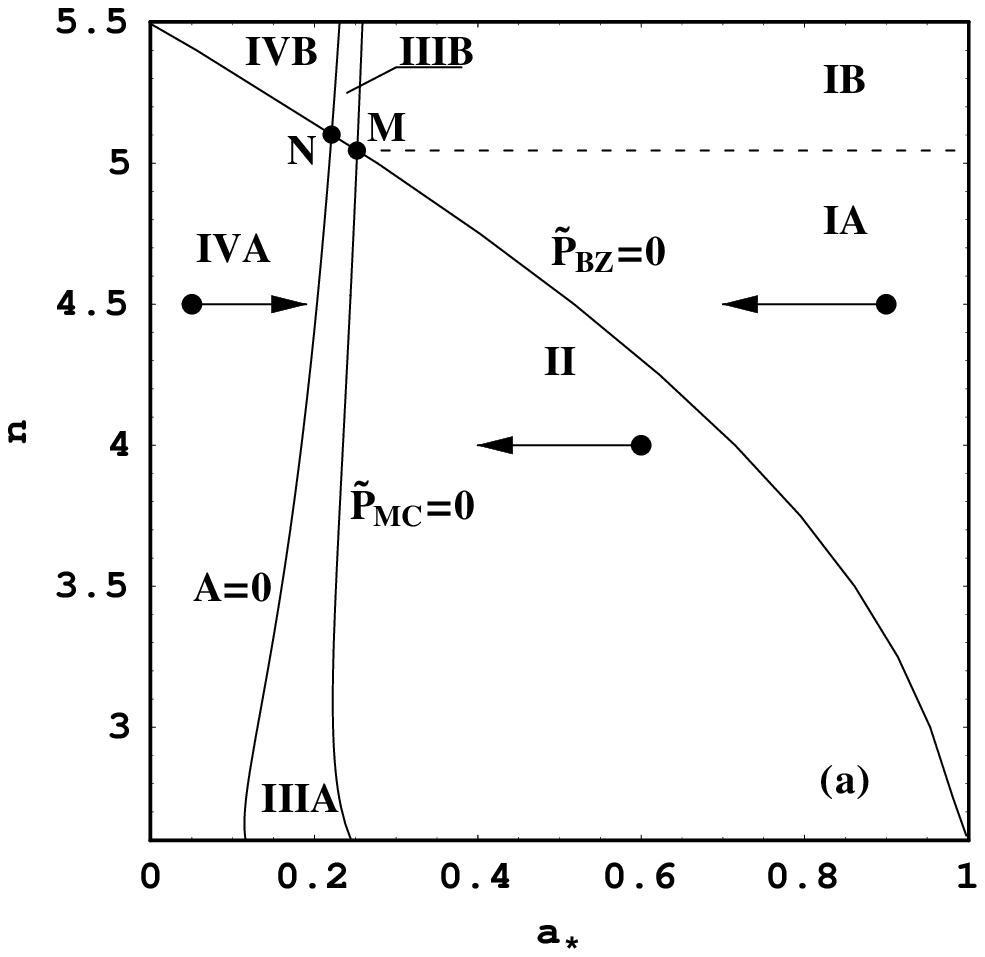}
 \end{center}
 \begin{center}
\epsscale{0.4}
 \plotone{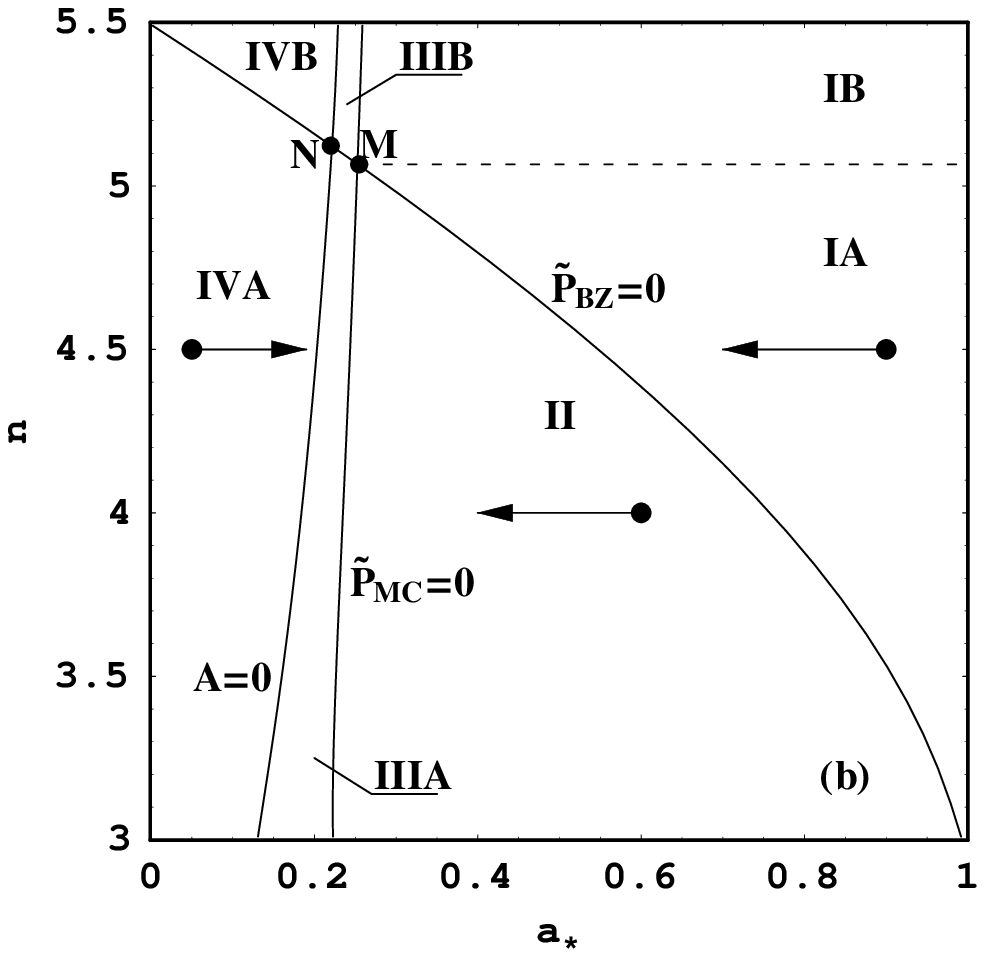}
 \caption{BH evolution is described in $a_ * - n$ parameter space with and
without screw instability in (a) and (b), respectively. The
parameter space is divided by the contours of $\tilde {P}_{BZ} = 0$,
that of $\tilde {P}_{MC} = 0$ and that of $A = 0$ into several
sub-regions. The duration of a GRB is regarded as the time for the
represent point of BH evolution moving from its initial position in
region IA to the contour $\tilde {P}_{BZ} = 0$. The dimensionless BZ
and MC powers are defined as $\tilde {P}_{BZ} \equiv P_{BZ} / P_0 $
and $\tilde {P}_{MC} \equiv P_{MC} / P_0 $, respectively. The
function $A = A(a_\ast ,n)$ is given in Lei05, being proportional to
the rate of change of the BH spin.}
 \label{fig3}
\end{center}
\end{figure}

\begin{figure}
\begin{center}
\epsscale{0.4}
 \plotone{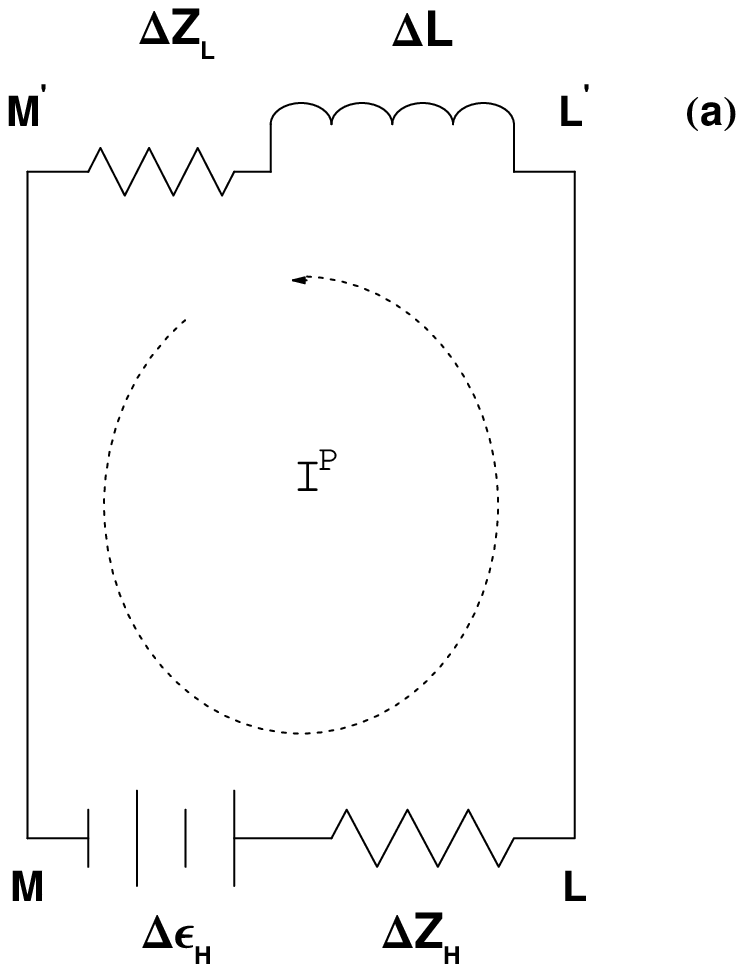}
 \end{center}
 \begin{center}
\epsscale{0.35}
 \plotone{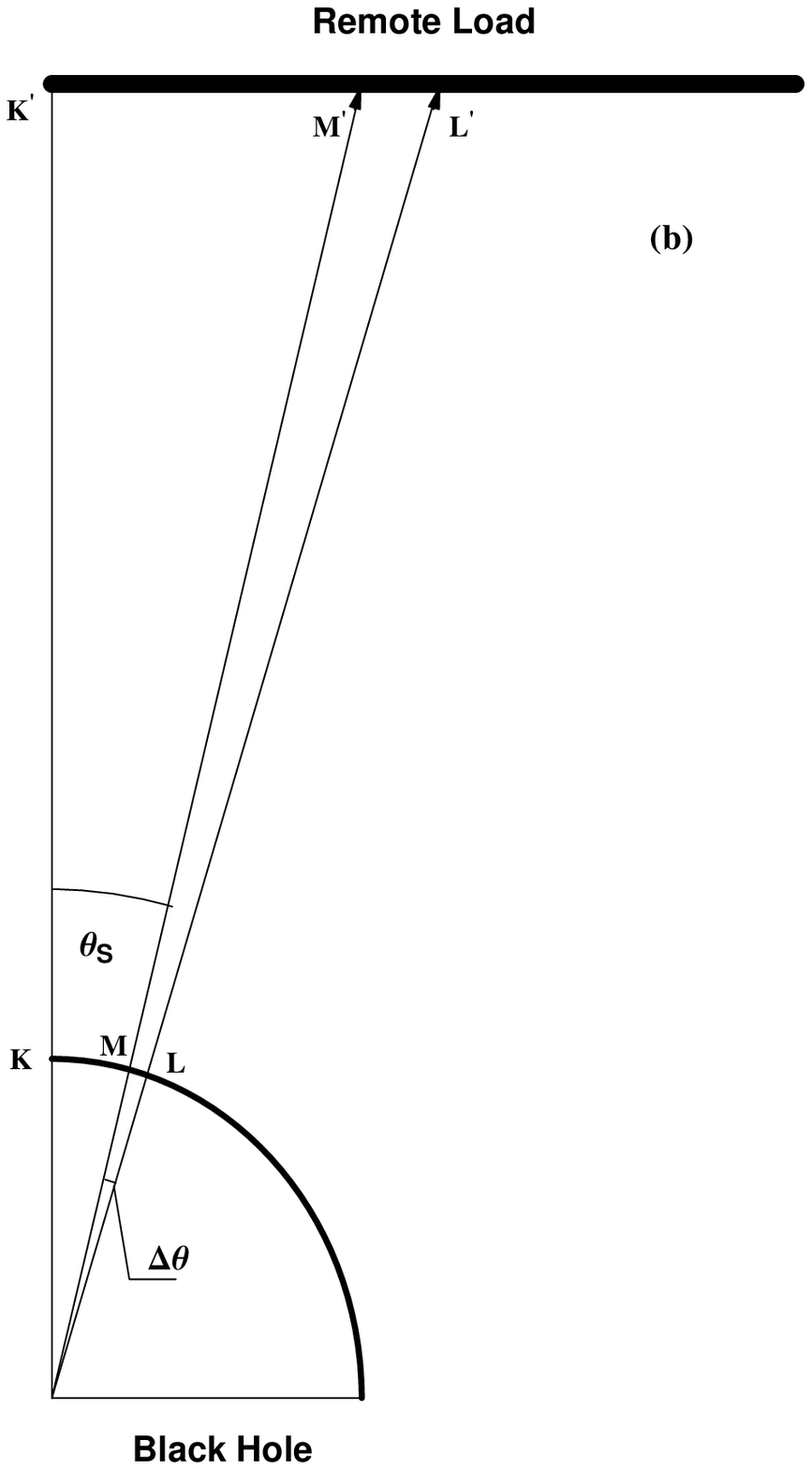}
 \caption{(a) An equivalent \textbf{\textit{R-L}}
circuit consisting of two magnetic surfaces, $M{M}'$ and $L{L}'$,
which connect the BH horizon and the remote load; (b) a simplified
configuration of poloidal magnetic field corresponding to the
equivalent \textbf{\textit{R-L}} circuit.}
 \label{fig4}
\end{center}
\end{figure}

\begin{center}
\begin{figure}
\epsscale{0.35}
\begin{center}
 \plotone{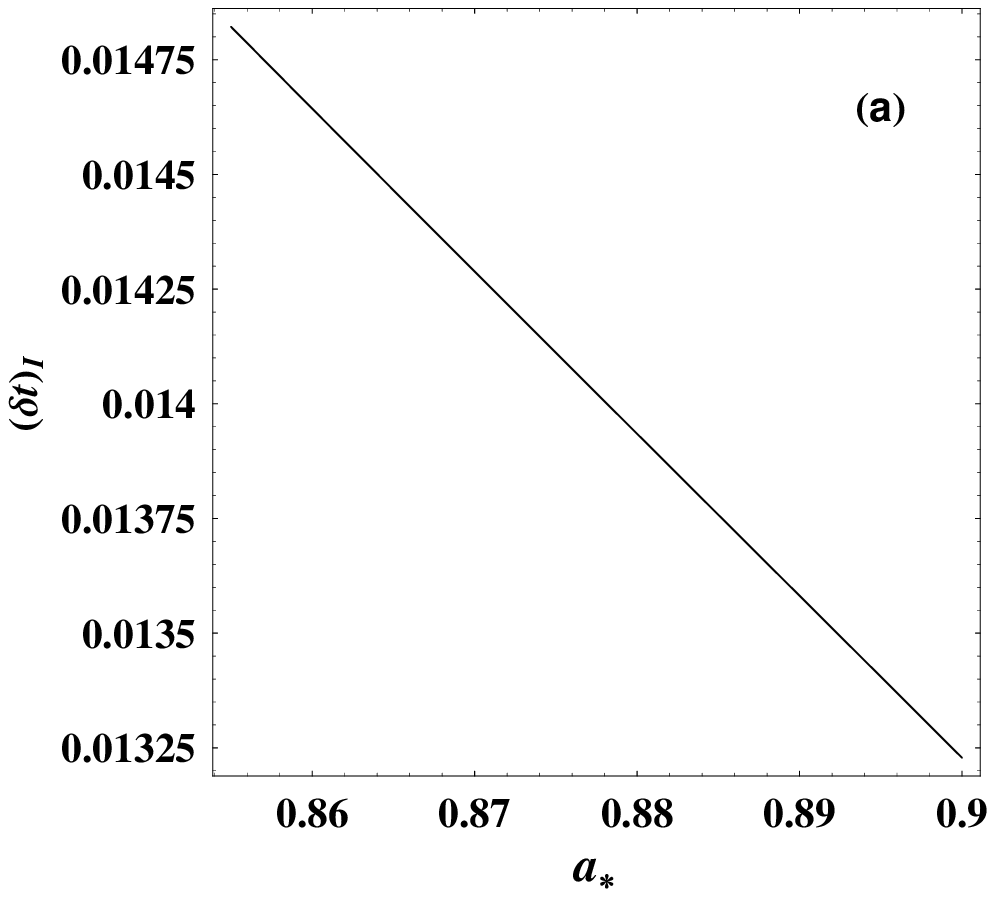}
\end{center}
\begin{center}
 \epsscale{0.35}
  \plotone{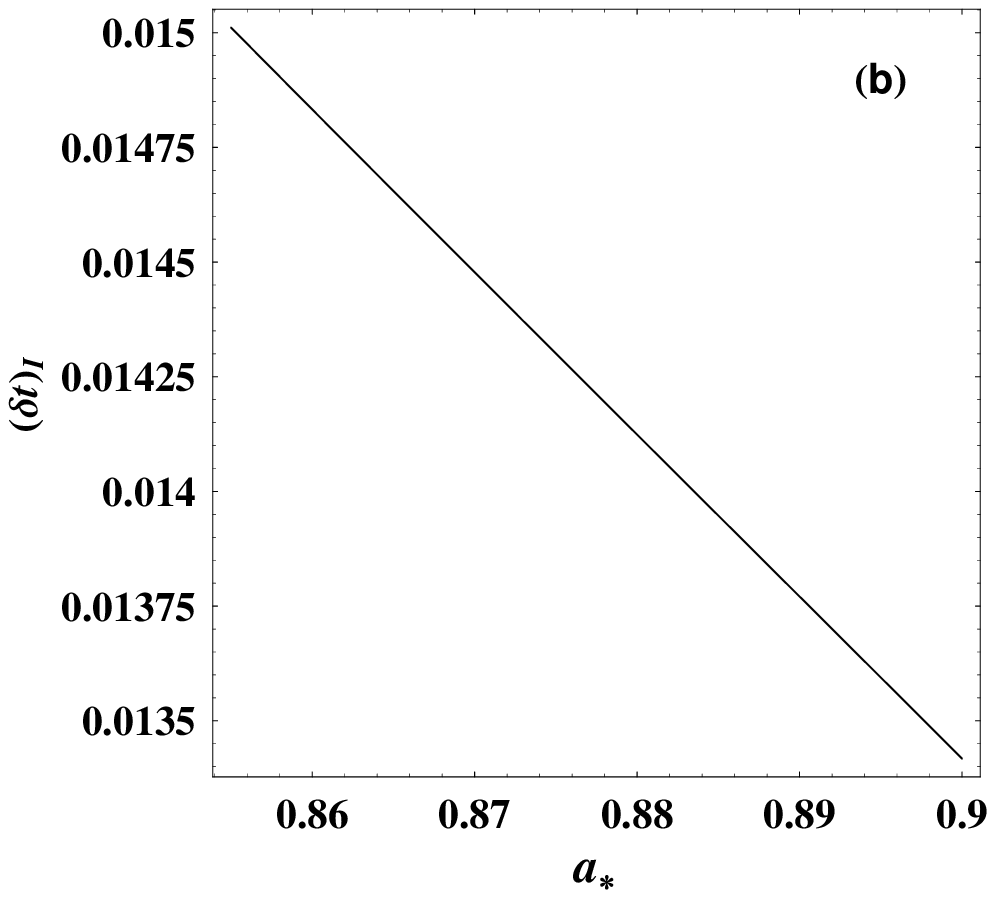}
\end{center}
\begin{center}
 \epsscale{0.35}
 \plotone{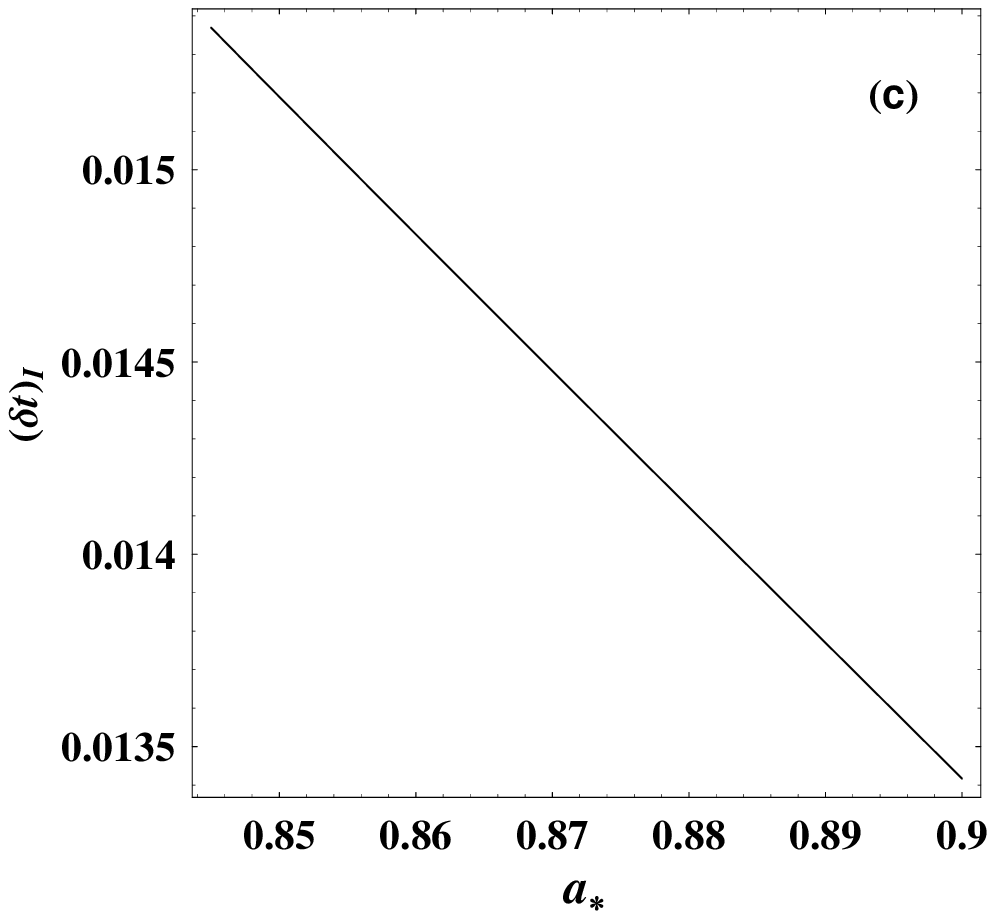}
 \end{center}
\caption{The curves of $\left( {\delta t} \right)_I $ versus $a_ * $
with n =3.54, 3.55 and 3.58 for GRBs 990712, 991208 and 991216 in
(a), (b) and (c), respectively.} \label{fig5}
\end{figure}
\end{center}


\clearpage

\begin{deluxetable}{ccccccc}
\tabletypesize{\scriptsize} \tablecaption{The Main Features of
Several Models Invoking the BZ Process For Powering GRBs}
\tablewidth{0pt} \tablehead{\colhead{Model} & \colhead{BZ Process}&
\colhead{MC Process} & \colhead{Surrounded By} & \colhead{Objects} &
\colhead{Half-Opening Angle} & \colhead{Variability} } \startdata

I & Yes& No& No disk& GRB& No&
No \\
\hline II& Yes& No& Transient \par Disk& GRB& No&
No \\
\hline III& Yes& No& Torus& GRB& No&
No \\
\hline IV& Yes& Yes& Disk& GRB/SN& No&
No \\
\hline V& Yes& Yes& Torus& GRB/SN& Yes&
No \\
\hline VI& Yes& Yes& Disk& GRB/SN& Yes&
No \\
\hline This Model& Yes& Yes& Disk& GRB/SN& Yes&
Yes \\
\hline
\enddata

\end{deluxetable}


\begin{deluxetable}{cccrrrrrr}
\tabletypesize{\scriptsize} \tablecaption{Five GRBs of True Energy
$E_\gamma $ and $T_{90} $s Fitted with Different Power-law Index $n$
and $B_H $ with the Predicted $E_{SN} $.} \tablewidth{0pt}
\tablehead{\colhead{GRB$^a$} & \colhead{$E_\gamma ^b$}
&\colhead{$T_{90} $} & \colhead{$n$} & &\colhead{$B_{15} ^h$}&
&\colhead{$E_{SN}^c$} & } \startdata

 970508& 0.234& 15$^d$& 3.89& 3.46& 0.97& 0.70& 1.947&
1.732 \\
\hline 990712& 0.445& 30$^e$& 3.98& 3.54& 0.81& 0.61& 1.989&
1.816 \\
\hline 991208& 0.455& 39.84$^f$& 3.99& 3.55& 0.75& 0.56& 1.995&
1.817 \\
\hline 991216& 0.695& 7.51$^f$& 4.06& 3.58& 1.80& 1.40& 2.032&
1.887 \\
\hline 021211& 0.117& 13.3$^g$& 3.81& 3.42& 0.9& 0.6& 1.901&
1.681 \\
\hline
\enddata

\tablecomments{\\ $^a$This sample of GRB-SNe is taken from Table 1
of Dar (2004). \\$^b$The true energies $E_\gamma $ of the GRBs are
given in terms of $10^{51}ergs$ from Table 1 of Frail et al (2001).
\\$^c$The predicted energies $E_{SN} $ of the SNe are given in terms
of $10^{51}ergs$ based our models. \\$^d$The duration of GRB 970508
is from Costa et al. (1997). \\$^e$The duration of GRB 990712 is
from Heise et al. (1999). \\$^f$The durations of GRBs 991208 and
991216 are from Table 2 of Lee {\&} Kim (2002).\\ $^g$The duration
of GRB 021211 is from Sakamoto et al. (2005). \\$^h$$B_{15} $ is the
magnetic field at the BH horizon in terms of $10^{15}gauss$. The
initial BH spin is assumed to be $a_\ast (0) = 0.9$, and the data in
the left and right sub-columns of the parameters $n$ and $B_{15} $
and $E_{SN} $ are calculated based on Model VI and this model,
respectively.}
\end{deluxetable}

\begin{deluxetable}{crrrrrrrr}
\tabletypesize{\scriptsize} \tablecaption{Five GRBs Fitted with
Different $f_{BZ} $ and $f_{MC} $ Invoking Model VI and This model.}
\tablewidth{0pt} \tablehead{\colhead{GRB} & \colhead{$a_\ast ^{GRB}
$} & &\colhead{$f_{BZ} $} & & \colhead{$f_{MC} $ }& &
\colhead{$a_\ast ^{SN} $}&  } \startdata

970508& 0.798& 0.871& 0.009& 0.020& 0.991& 0.980& 0.233&
0.226 \\
\hline 990712& 0.767& 0.853& 0.014& 0.028& 0.986& 0.972& 0.234&
0.227 \\
\hline 991208& 0.763& 0.852& 0.014& 0.029& 0.986& 0.971& 0.235&
0.227 \\
\hline 991216& 0.736& 0.842& 0.019& 0.032& 0.981& 0.968& 0.236&
0.228 \\
\hline 021211& 0.823& 0.880& 0.006& 0.016& 0.994& 0.984& 0.232&
0.225 \\
\hline

\enddata
\end{deluxetable}


\begin{deluxetable}{ccccccccc}
\tabletypesize{\scriptsize} \tablecaption{Variability Time Scale in
the Light-Curves of Five GRBs.} \tablewidth{0pt}
\tablehead{\colhead{GRBs} &
 &\colhead{Parameters} & & &\colhead{$T_{90} $(s)} & &
\colhead{Variablity Time scale (ms)}&  } \startdata
 &
$a_ * $& $n$& $\tilde {r}_S $& $h_c $&
 &
$\left( {\delta t} \right)_I $& $\left( {\delta t} \right)_{II} $&
$\left( {\delta t} \right)_{III} $ \\
\cline{2-5} \cline{7-9} 970508& 0.9& 3.46& 5.96& 87.71& 15& 13.42&
26.84&
40.25 \\
\hline 990712& 0.9& 3.54& 6.45& 85.56& 30& 13.23& 26.46&
39.68 \\
\hline 991208& 0.9& 3.55& 6.52& 87.71& 39.84& 13.42& 26.84&
40.25 \\
\hline 991216& 0.9& 3.58& 6.77& 87.70& 7.51& 13.42& 26.83&
40.25 \\
\hline 021211& 0.9& 3.42& 5.77& 59.32& 13.3& 10.90& 21.80&
32.70 \\
\hline

\enddata

\tablecomments{\\ $\tilde {r}_S \equiv {r_{_S} } \mathord{\left/
{\vphantom {{r_{_S} } M}} \right. \kern-\nulldelimiterspace} M$ is
the critical dimensionless radius for SIMC.}

\end{deluxetable}

\end{document}